\documentclass{article}
\usepackage[english]{babel}
\usepackage[T1]{fontenc}
\usepackage{fullpage}
\usepackage[utf8]{inputenc}
\usepackage{mathrsfs} 
\usepackage{newlfont}
\usepackage{fancyhdr}
\usepackage{mhchem}
\usepackage{graphicx}
\usepackage{newlfont}
\usepackage{amssymb}
\usepackage{amsmath}
\usepackage{latexsym}
\usepackage[normalem]{ulem}
\usepackage{amsthm}
\usepackage{bm}
\usepackage{bbold}
\usepackage{xcolor}
\usepackage{cancel}
\usepackage{listings}
\usepackage[hidelinks]{hyperref}
\usepackage[capitalize]{cleveref}
\usepackage{subfigure}
\usepackage{dsfont}
\hypersetup{
    colorlinks=true,
    linkcolor=black,
    citecolor=blue,
    filecolor=black,
    urlcolor=blue,
    pdfauthor={M. Cadoni, M. Oi, M. Pitzalis, A. P. Sanna},
    pdftitle={Scalar stars and lumps with (A)dS core}
}
\usepackage{cite}

\newcommand{\rH}{r_\text{H}}

\newcommand{\dd}{\text{d}}

\title{\bf Scalar stars and lumps with (A)dS core}
\author{M.~Cadoni${}^{ab}$\thanks{E-mail: mariano.cadoni@ca.infn.it}, \, M.~Oi${}^{ab}$\thanks{E-mail: mauro.oi@ca.infn.it},\, M. Pitzalis${}^{ab}$\thanks{E-mail: mirko.pitzalis@ca.infn.it}, \, A.~P.~Sanna${}^{ab}$\thanks{E-mail: asanna@dsf.unica.it} \ 
\\
${}^a$\emph{Dipartimento di Fisica, Universit\`a di Cagliari}
\\
{\em Cittadella Universitaria, 09042 Monserrato, Italy}
\\
\\
${}^b$\emph{I.N.F.N, Sezione di Cagliari}
{\em  Cittadella Universitaria, 09042 Monserrato, Italy}
\\}
\begin{document}
\maketitle
\begin{abstract}
We explore the possibility of embedding regular compact objects with (anti) de Sitter ((A)dS) core as solutions of Einstein's gravity minimally coupled to a real scalar field. We consider, among others,  solutions interpolating between an inner, potential-dominated core and an outer, kinetic-term-dominated region. Owing to their  analogy with slow-roll inflation, we term them gravitational vacuum inflative stars, or gravistars for short. We systematically discuss approximate solutions of the theory describing either the core or the asymptotically-flat region at spatial infinity. We extend nonexistence theorems for smooth interpolating solutions, previously proved for black holes, to compact objects without event horizons.  This allows us to construct different classes of exact (either smooth or non-smooth) singularity-free solutions of the theory. We first find a smooth solution interpolating between an AdS spacetime in the core and an asymptotically-flat spacetime (a Schwarzschild solution with a subleading $1/r^2$ deformation). We proceed by constructing non-smooth solutions describing gravistars. Finally, we derive a smooth scalar lump solution interpolating between $\text{AdS}_4$ in the core and a Nariai spacetime at spatial infinity. 
\end{abstract}

\newpage
\tableofcontents
\section{Introduction and motivations}

To date, a considerable amount of observational evidence \cite{LIGOScientific:2016aoc, LIGOScientific:2017vwq,EventHorizonTelescope:2019dse,EventHorizonTelescope:2022wkp,Eckart:1997em} has accumulated confirming the existence of one of the most enigmatic solutions of general relativity (GR), black holes.
Nevertheless, these objects are still mysterious, as they entail profound puzzles that challenge our current understanding of gravitational physics. First of all, behind their event horizons, they harbor an unavoidable singularity \cite{Penrose:1964wq}, which is a byproduct of the stellar collapse leading to their formation. This singularity marks a point where GR loses its predictive power entirely. Second of all, the information paradox related to their evaporation \cite{Hawking:1976ra,Page:1993wv,Mathur:2009hf} still embodies the most formidable obstacle to reconciling gravity with quantum mechanics. 

These challenges have sparked, over the years, a plethora of alternatives to classical singular black holes, either as solutions of theories beyond GR or as \emph{ad-hoc} effective models (for an incomplete list, see, e.g., Refs. \cite{Mathur:2005zp,Cadoni:2015gfa,Herdeiro:2015waa,Frolov:2016pav,Cardoso:2016rao,Franzin:2018pwz,Carballo-Rubio:2018jzw,Cardoso:2019rvt,Simpson:2019mud,Carballo-Rubio:2019fnb,Brustein:2019bou,Lobo:2020ffi,Ikeda:2021uvc,Akil:2022coa,Blazquez-Salcedo:2022dxh,Dymnikova:1992ux,Bronnikov:2005gm,Mazza:2021rgq,Simpson:2021dyo,Xu:2021lff,Sebastiani:2022wbz,Berry:2021hos,Carballo-Rubio:2022nuj,Rodrigues:2022qdp,Bronnikov:2022ofk,Carballo-Rubio:2023mvr,Bertipagani:2020awe,Borissova:2022mgd,Mazza:2023iwv,Bronnikov:2022gjq,Bonanno:2020fgp,Bonanno:2000ep,Platania:2019kyx,Mazur:2001fv,Visser:2003ge,Mazur:2004fk,MartinMoruno:2011rm,Lobo:2010uc} and references therein), which act as black-hole mimickers. Recent decades have witnessed a resurgence in this field, driven by cutting-edge experiments that have the potential to detect deviations from standard classical GR predictions. Indeed, despite the observational evidence being largely consistent with the presence of Kerr black holes \cite{LIGOScientific:2016aoc,EventHorizonTelescope:2019dse,EventHorizonTelescope:2022wkp,Eckart:1997em,Ghez:2000ay,Bambi:2011mj,Bambi:2016sac,Cardenas-Avendano:2016zml}, there is still room for slight deviations from GR that could be tested by future experiments since they could provide some evidence for the existence of the aforementioned mimickers. 

Among the latter, some peculiar solutions describing asymptotically-flat compact object with a de Sitter (dS) core seem particularly promising. Two paradigmatic examples are the well-known gravastars \cite{Mazur:2001fv,Visser:2003ge,Mazur:2004fk,DeBenedictis:2005vp,Pani:2009ss,Pani:2010em,Mottola:2010gp,Lobo:2010uc,MartinMoruno:2011rm,Mottola:2022tcn,Mottola:2023jxl} and specific models of singularity-free black holes/compact objects (for an incomplete list, see Refs.~\cite{bardeen1968proceedings,Dymnikova:1992ux,Bonanno:2000ep,Hayward:2005gi,Ansoldi:2008jw,Spallucci:2011rn,Modesto:2008im,Nicolini:2008aj,Fan:2016hvf,Cadoni:2022chn}). These are exact solutions of Einstein's equations sourced by an anisotropic fluid (see, \emph{e.g.}, Ref.~\cite{Cadoni:2022chn} and references therein), with the peculiar equation of state (EoS) $p_\parallel = -\rho$ (where $\rho$ is the energy density of the fluid, while $p_\parallel$ its radial pressure). The latter resembles the dark-energy EoS and is able to generate the dS core, whose advantage is twofold. It breaks the strong energy condition, allowing therefore to circumvent Penrose's theorem and to construct nonsingular solutions. Additionally, it suggests an intriguing connection with our current accelerated expansion of the universe, compatible with the dominance of a positive cosmological constant \cite{SupernovaSearchTeam:1998fmf,SupernovaSearchTeam:2004lze,Planck:2018vyg}. To realize the interpolation between the dS core and an asymptotically-flat spacetime, an additional length-scale parameter $\ell$ must be included in the solution. This represents an additional ``hair'', which can be interpreted \cite{Cadoni:2022chn} as encoding quantum-gravity corrections responsible for the smearing of the singularity. Depending on the value of $\ell$ with respect to the classical Schwarzschild radius $R_\text{S} = 2 G M$, these models can: $(1)$ have two horizons, if $\ell \ll R_\text{S}$; $(2)$ be extremal when $\ell \sim R_\text{S}$, when the two horizons degenerate into a single one; $(3)$ be horizonless objects when $\ell > R_\text{S}$. Intriguingly, both theoretical \cite{Cadoni:2022chn,Cadoni:2023tse} and observational \cite{Lamy:2018zvj,Guo:2021bhr,Riaz:2022rlx,Cadoni:2022vsn,Boshkayev:2023rhr} evidence suggests that models with quantum-gravity ``superplanckian'' corrections acting at the Schwarzschild radius could be particularly relevant.

However, the intrinsic effective nature of these models, which are still rooted in GR, poses difficulties in the interpretation of the ``hair'' $\ell$. The microscopic origin of the latter is, indeed, still poorly understood\footnote{The only known ``microscopic'' interpretation one can resort to is a magnetic charge in solutions sourced by nonlinear electrodynamics \cite{Ayon-Beato:1998hmi,Bronnikov:2000vy,Dymnikova:2004zc}.}. 
This difficulty becomes even more severe when the deformations from the usual Schwarzschild solution have superplanckian origin. This leads to a lack of predictability, as there is an infinite degeneration of regular models with the same qualitative properties, some of which also arise as solutions of different quantum-gravity theories \cite{Bonanno:2000ep,Modesto:2008im,Nicolini:2008aj,Nicolini:2019irw}. 

In light of this situation, there is a clear need to build a unified framework capable of including these models as solutions, but, at the same time, providing the predictability needed to distinguish between physical and unphysical solutions. An interesting possibility would be a scalar-field theory coupled to gravity. This theory is the simplest to describe an anisotropic fluid and, therefore, an excellent candidate to transition from a coarse-grained, fluid-based description to an elementary one in terms of fields. Scalar fields are also pervasive in theoretical physics, offering solutions that interpolate between different vacua (Minkowski, dS, AdS) \cite{Bronnikov:2005gm,Bronnikov:2022gjq,Cadoni:2009xm,Cadoni:2012uf,Luo:2022roz,Gregory:2018bdt,Li:2022ayz,Moss:1984zf,Karakasis:2023hni,Bakopoulos:2023hkh}, making them potentially capable of realizing the seeked interpolation between the dS core and an asymptotically-flat spacetime. Additionally, scalar models describe transitions between different spacetimes, such as AdS and dS \cite{Biasi:2022ktq,Garriga:2013cix}, and are profusely employed in the AdS/CFT correspondence \cite{Hartnoll:2008vx,Hartnoll:2008kx,Horowitz:2008bn,Cadoni:2011kv}. Finally, cosmological observations align with a primordial inflationary era driven by a scalar field \cite{Guth:1980zm,Linde:1981mu,Liddle:1994dx,Guth:2007ng,Linde:2007fr,Senatore:2016aui,Planck:2018jri,Achucarro:2022qrl}, akin to the current accelerated expansion of the universe.

The existence of (singular) black-hole solutions in Einstein-scalar gravity is a topic which has been widely investigated in the past, particularly in connection with the possibility of providing the GR Schwarzschild solution with some scalar ``hair'' \cite{Israel:1967wq,Bekenstein:1995un}. The main outcome of these investigations has been the formulation of no-hair theorems forbidding the existence of such ``hairy'' black holes and the possibility of circumventing them by choosing particular forms of the scalar-field potential \cite{Cadoni:2015gfa,Herdeiro:2014goa,Franzin:2018pwz}. Such investigations have also been extended to the case of regular black-hole solutions, i.e., solutions with horizons, but without a central singularity \cite{Bronnikov:2001ah,Bronnikov:2001tv,Galtsov:2000wmw}. 

Since, up to now, very little is known about regular, horizonless, static compact solutions of Einstein-scalar gravity, the main purpose of this paper is to investigate in a systematic way the existence of such solutions.  
One of the main complications when dealing with solutions of Einstein-scalar gravity is that, in the usual approach, one fixes the form of the scalar-field potential from the beginning, restricting drastically the landscape of possible solution.
In this paper, we will go around this problem by using a different parametrization of the solutions, which uses the radial metric function instead of the potential.
We will try to understand if a compact horizonless objects, whose geometry smoothly interpolates between a dS core and an asymptotically-flat spacetime, is allowed in such a theory. Although previous no-go theorems \cite{Bronnikov:2001ah,Bronnikov:2001tv,Galtsov:2000wmw} have ruled out this possibility for compact objects with horizons, the same is not true for horizonless compact objects. The present work extends those theorems to encompass the latter, relying solely on the dynamical equations. Understanding these no-go theorems will open the way to circumvent them. We will construct different classes of exact (smooth and non-smooth) singularity-free solutions of the theory, describing stars with a dS and AdS core or scalar lumps interpolating between different vacua of the theory. 

The structure of the paper is the following. \\
In \cref{sec:GRscalars} we briefly review the main properties of minimally coupled Einstein-scalar gravity, its vacuum solutions and the solution-generating algorithm, firstly employed in Refs. \cite{Franzin:2018pwz,Cadoni:2018pav} (see also \cite{Cadoni:2011nq,Cadoni:2015gfa}). We also propose solutions interpolating between an inner core region, dominated by the potential, and an outer one, dominated instead by the scalar-field kinetic term. Given the analogy with the slow-roll mechanism in inflationary cosmology, we term these gravitational vacuum inflative stars, or gravistars for short. \\
In \cref{sec:asympsols}, we study approximate solutions, both in the form of a dS spacetime in the inner core and of asymptotically-flat solutions, with a leading Schwarzschild behavior, at spatial infinity.\\
In \cref{sec:NoGodS} we prove a nonexistence theorem for smooth gravistar solutions.\\
In \cref{sec:smoothAdScore,sec:JointedSol,sec:ScalarLump} we construct solutions, which circumvent our nonexistence theorem. By trading the dS core for an AdS one, we construct smooth star solutions interpolating between the AdS and the Schwarzschild spacetimes (\cref{sec:smoothAdScore}). By giving up smoothness, we construct jointed gravistar solutions (\cref{sec:JointedSol}). Finally, by giving up asymptotic flatness, in \cref{sec:ScalarLump} we derive scalar lump solutions which interpolate between the $\text{AdS}_4$ spacetime in the interior and a Nariai spacetime at infinity.\\
We draw our conclusions in \cref{sec:conclusions}. 

\section{General Relativity minimally coupled with a scalar field}
\label{sec:GRscalars}

Our starting point is Einstein's gravity minimally coupled to a real scalar field $\phi$ \footnote{Throughout this work, otherwise explicitly stated, we adopt units in which $c=\hbar = 16\pi G = 1$. }
\begin{equation}\label{action}
\mathcal{S} = \int \dd^4 x \, \sqrt{-g} \left(\mathscr{R}-\frac{1}{2}g^{\mu\nu}\partial_\mu \phi \partial_\nu \phi - V(\phi) \right)
\end{equation}
where $\mathscr{R}$ is the Ricci scalar, while $V(\phi)$ is the self-interaction potential.
We look for asymptotically-flat, spherically-symmetric, static solutions of the field equations stemming from the action \eqref{action}, of the form $\dd s^2 = -U(r) \dd t^2 + U(r)^{-1} \dd r^2 + R(r)^2 \dd \Omega^2$, with $\phi= \phi(r)$.
It has been shown that this kind of solutions can be fully parametrized in terms of a single function $P(x)$ of the dimensionless spacelike coordinate $x \equiv r_0/r$ ($r_0$ is an arbitrary length scale)~\cite{Franzin:2018pwz}. 
We have
\begin{subequations}
\begin{align}
&R(x) = \frac{r_0}{x}P\, ; \label{Rgeneral}\\
&\phi(x) = 2\int \dd x \, \sqrt{-\frac{1}{P}\frac{\dd^2 P}{\dd x^2}}\, ; \label{phigeneral}\\
&U(x) = \frac{r_0^2 P^2}{x^2}\left(c_2 + \frac{2}{r_0^2}\int \frac{\dd x \, x}{P^4} + \frac{c_1}{r_0^3}\int \frac{\dd x \, x^2}{P^4}\right)\, ; \label{Ugeneral}\\
&V\left[\phi(x)\right] = \frac{x^2}{2r_0^2 P^2}\left[2-x^2 \frac{\dd}{\dd x}\left(x^2 \frac{\dd}{\dd x} \frac{U P^2}{x^2} \right) \right]\, , \label{Vgeneral}
\end{align}
\label{solutiongeneral}
\end{subequations}
where $c_{1,2}$ are two integration constants, whose values are determined by boundary conditions. 
With the previous parametrization we can have different, equivalent branches of the solutions depending on the sign of $x$ and $P$. In this paper we will consider the branch for which $x \ge 0$ and $P \ge 0$. Note that asymptotic infinity $r \to \infty$ corresponds to $x\to 0$, while the region near $r \to 0$ corresponds to $x \to \infty$.
Until now, the parametrization \eqref{solutiongeneral} has been mainly used to set up a solution-generating algorithm, providing the form of the potential $V$ once either a particular radial function or a scalar-field profile is assumed \cite{Franzin:2018pwz,Cadoni:2018pav,Cadoni:2011nq}.
In the following, we will use it in a systematic way to investigate asymptotic solutions of the theory and to formulate a nonexistence theorem.

\subsection{Vacuum solutions}
\label{GRCoupledScalarFieldExactSolution}

When the potential $V$ is constant or has a local extremum, the system \eqref{solutiongeneral} has solutions with constant $\phi=\phi_0$, 
corresponding to pure GR. Depending on the sign of $V(\phi_0) $, we have either the standard Schwarzschild, dS  or AdS   solutions for $V(\phi_0)=0$, $V(\phi_0)>0$ or $V(\phi_0)<0$, respectively. 
In terms of the parametrization used in \cref{solutiongeneral}, these solutions are given by 
\begin{equation}\label{vsol}
P(x)=1, \qquad  U_\text{GR}(x) = 1 + \frac{c_1}{3 r_0}x + \frac{c_2 r_0^2}{x^2}, \qquad \phi(x)=\phi_0, \qquad  V(x) = -6 c_2\, .
\end{equation}
The linear term in \cref{vsol} gives the usual Schwarzschild contribution. Therefore, $c_1$ is given in terms of the ADM mass $M$ of the solution by $c_1 = -3M/8\pi$ \footnote{This holds generically for all the solutions we consider in this paper. In fact, the flat or (A)dS asymptotic at $r\to\infty$ requires $P \sim 1$ ($R \sim r$), so that the third term in \cref{Ugeneral}, once integrated, gives a term proportional to $x \sim 1/r$.}. The last term is the usual (A)dS term. In particular, asymptotically we have either Minkowski, dS or AdS vacua for $c_2=0$, $c_2<0$ or $c_2>0$, respectively, which are also the only solutions we will consider at the core ($x\to \infty$) and at asymptotic infinity ($x \to 0$).

\subsection{Gravitational vacuum inflative stars (Gravistars)}
\label{subsec:inflstars}

Solutions of Einstein's gravity describing asymptotically-flat compact objects with a dS core have recently received renewed and increasing interest \cite{Cadoni:2022chn,Brustein:2019bou,Lemos:2011dq,Dymnikova:2016nlb,Dymnikova:2017pos,Sebastiani:2022wbz}. Two paradigmatic examples are the well-known gravastars \cite{Mazur:2001fv,Visser:2003ge,Mazur:2004fk,DeBenedictis:2005vp,Pani:2009ss,Pani:2010em,Mottola:2010gp,MartinMoruno:2011rm,Mottola:2022tcn,Mottola:2023jxl} and specific models of singularity-free black holes. These are exact solutions of Einstein's equations sourced by an anisotropic fluid (see, \emph{e.g.}, Refs.~\cite{Cadoni:2022chn,Sebastiani:2022wbz} and references therein)

\begin{equation}
    T_{\mu\nu} = \left(\rho + p_\perp \right)u_\mu u_\nu + p_\perp g_{\mu\nu} + \left(p_\parallel-p_\perp \right)w_\mu w_\nu\, ,
    \label{AnistropicSEI}
\end{equation}
where $\rho$, $p_\parallel$ and $p_\perp$ are the energy density, the radial and perpendicular components of the pressure, respectively, while $u_\mu$ and $w_\mu$ are a timelike and a spacelike 4-vector, respectively, satisfying the normalization conditions $u^\mu u_\mu = -w^\mu w_\mu = -1$. In particular,  this form of the stress-energy tensor allows to circumvent standard singularity theorems \cite{Penrose:1964wq} and to construct singularity-free black-hole solutions. The EoS characterizing the latter is typically of the form $p_\parallel = -\rho$.

A drawback of the approach based on the generic anisotropic fluid as a source is the lack of tight constraints on the form of the functions parametrizing the geometry of these regular models (see the discussions in Refs.~\cite{Cadoni:2022chn,Cadoni:2023nrm}). 
An interesting possibility, which could improve the situation, would be to consider the abovementioned models as  solutions of gravity minimally coupled with a scalar field. Indeed, the latter is the simplest theory characterized by an anisotropic stress-energy tensor, which reads as
\begin{equation}\label{SETscalar}
T^{(\phi)}_{\mu\nu} = \partial_\mu \phi \partial_\nu \phi - g_{\mu\nu} \left(\frac{1}{2} g^{\alpha \beta}\partial_\alpha \phi \partial_\beta \phi + V(\phi) \right)\, .
\end{equation}
However, it is easy to show that \cref{SETscalar} is, in general, incompatible with the required $p_\parallel = -\rho$ EoS. In fact, from the component of the stress-energy tensor we have
\begin{subequations}
\begin{align}
&T^{(\phi) 0}_0 = -\rho = -\frac{U}{2}\phi'^2 - V\, ;\\
&T^{(\phi) r}_r  = p_\parallel = \frac{U}{2} \phi'^2 -V\, ;\\
&T^{(\phi) \theta}_\theta = p_\perp = -\frac{U}{2}\phi'^2 -V\, ,
\end{align}
\label{EnergyMomentumComponents}
\end{subequations}
where the prime refers to derivation with respect to $r$. Therefore, regular models characterized by the EoS $p_\parallel = -\rho$ cannot be obtained as \emph{exact} solutions of Einstein's gravity coupled to a scalar field. 

Given this, we can still hope to generate approximate solutions to mimic the behavior of the compact objects discussed above. This can be done by requiring that the EoS $p_\parallel = -\rho$ is satisfied in the core, where it must source the dS spacetime, and by letting the EoS deviate from this form at different locations. Moreover, when the solution is asymptotically flat, the fluid must again satisfy the EoS $p_\parallel = -\rho$, but with $\rho=0$. 

Using \cref{EnergyMomentumComponents}, the conditions above can be reformulated in terms of inequalities involving the ``kinetic energy'' $T=U\phi'^2 /2$ and the potential $V$ of the scalar field. In particular, $V$ must dominate over $T$ in the core, \emph{i.e.}, at $r \to 0$, where $V \gg T$, whereas both $V$ and $T$ need to asymptote $0$ for $r \to \infty$. However, the last equation does not specify whether the potential or the kinetic term dominates at spatial infinity. A hint to settle this issue comes from the investigations on the scale symmetries characterizing nonsingular black holes with a dS core \cite{Cadoni:2023nrm}. It has been shown that, in the most relevant physical case, a scale symmetry emerges in the asymptotic region \cite{Cadoni:2023nrm}. This motivates us to select models characterized by $T \gg V$ at $r \to \infty$, allowing the scale symmetry, typical for conformal field theories, to emerge.

The scenario just described somewhat resembles the inflationary one in cosmology \cite{Guth:1980zm,Linde:1981mu,Liddle:1994dx,Guth:2007ng,Linde:2007fr,Senatore:2016aui,Achucarro:2022qrl}, where the time coordinate is, here, replaced by the radial one.
Since, in the following, we will investigate models interpolating from a potential-dominated regime near $r\to0$ (dS/AdS vacuum) to a kinetic-term-dominated regime near $r\to \infty$ (Minkowski vacuum), we will call these (possible) solutions of Einstein-scalar gravity gravitational vacuum inflative stars, or \emph{gravistars} for short. In the following three sections, we will investigate the existence of these solutions.
  
\section{Asymptotic solutions}
\label{sec:asympsols}

Let us first derive approximate solutions of the system \eqref{solutiongeneral} in the two asymptotic regions $r\to0$, \emph{i.e.}, $x\to \infty$ (the core), and $r\to \infty$, \emph{i.e.}, $x\to0$ (spatial infinity).

\subsection{Approximate solution in the core}
\label{subsec:Aroundzero}

We assume that our solution has a leading dS behavior near $x \to \infty$ by requiring the leading behaviour
\begin{equation}\label{asmetric}
P=1, \qquad U= 1- \left(\frac{r_0}{x \hat L }\right)^2\, ,
\end{equation}
and then we derive the subleading terms allowed by the dynamical equations \eqref{solutiongeneral}. Here $\hat L$ is the dS length.

The approximate solutions can be classified by the form of the allowed subleading terms in $U$. We limit our discussion to the first relevant subleading terms, after the quadratic one, up to order $\mathcal{O}(x^{-5})$. Apart from the simplicity, this choice is also motivated  by the behavior of relevant well-known, asymptotically-flat, regular black-hole solutions with a dS core. Some notable examples are given below in terms of the radial coordinate $r$ \footnote{In the present section, we reinstate $G$ to make the contact with these solutions easier.}

\begin{align}
\nonumber& \text{Fan\& Wang} &U_\text{FW}(r) &= 1-\frac{2GM r^2}{(r+\ell)^3} \simeq 1-\frac{2GM r^2}{\ell^3} + \frac{6GM r^3}{\ell^4} + \mathcal{O}(r^{4})\, ,\\
\label{regularmetricr0}
& \text{Bardeen} &U_\text{B}(r) &= 1-\frac{2GM r^2}{(r^2 + \ell^2)^{3/2}} \simeq 1-\frac{2GM r^2}{\ell^3} + \frac{3GM r^4}{\ell^5} + \mathcal{O}(r^{5})\, ,\\
\nonumber& \text{Hayward} &U_\text{H}(r) &= 1-\frac{2GM r^2}{r^3 + \ell^3} \simeq 1-\frac{2GM r^2}{\ell^3} + \frac{2GM r^5}{\ell^6} + \mathcal{O}(r^{6})\, ,
\end{align}
where, in all the above, $\ell$ is a length scale responsible for the smearing of the classical singularity and parametrizing, together with the classical Schwarzschild radius $2GM$, the dS length $\hat L$ (see Ref.~\cite{Cadoni:2022chn}).
It is worth noticing that the leading and first subleading terms, giving the dS behavior near the core, are universal, while the corrections are model-dependent, as they determine the junction with the behavior at asymptotic infinity. These subleading corrections are all positive to guarantee this interpolation, which in turn determines the presence of a minimum. 

Now we exploit the solution generating algorithm, given by \cref{solutiongeneral}, adopting the following expansion of $P(x)$ around $x \to \infty$ 
\begin{equation}\label{eqp}
P(x) = 1+  \frac{a_1}{r_0 x}+ \frac{a_2}{r_0 x^2} + \frac{a_3}{r_0 x^3} + \frac{a_4}{r_0x^4} + \frac{a_5}{r_0x^5} + ...
\end{equation}
where the $a_i$'s are constants with dimensions of length. 
One can easily check, using the field equations \eqref{solutiongeneral}, that the second and third terms in the previous equation produce, respectively, a linear and logarithmic term in $U(x)$, which are incompatible with the leading term in \cref{regularmetricr0} and require, therefore, $a_1=a_2=0$.

On the other hand, the fourth term produces a subleading contribution which scales as the Fan $\&$ Wang metric in \cref{regularmetricr0}. For $a_1=a_2=0$, indeed, the function $U$ behaves as 
\begin{equation}\label{eqforu}
U(x) \simeq 1 - \frac{r_0^2}{\hat L ^2 x^2} + \frac{10 a_3}{r_0 \, x^3} + \frac{6a_4}{r_0 \, x^4} + \mathcal{O}(x^{-5})\, ,
\end{equation}
with $1/\hat L^2= -c_2$, when $a_4\neq 0$, whereas $1/\hat L^2= -c_2 + 220 \pi \, a_3^{2/3}/81 \sqrt{3} \, r_0^{8/3}$, when $a_4=0$. In order to have an asymptotic solution consistent with a dS core (see \cref{GRCoupledScalarFieldExactSolution}), $c_2<0$ is required\footnote{If $c_2 >0$, the solution is characterized by an AdS core instead, and the signs in both the definition of $\hat L$ and in the second term of \cref{eqforu} are opposite.}.
The potential $V$ and the scalar field $\phi$ behave as
\begin{equation}\label{potphi}
V(x) \simeq \frac{6}{\hat L^2}  - \frac{120 a_3}{r_0^3 \, x} + \mathcal{O}(x^{-2}), \qquad \phi(x) \simeq  4 \sqrt{-\frac{a_3}{3r_0}}\, x^{-3/2} + \mathcal{O}(x^{-2})\, .
\end{equation}
Notice that the condition $a_3<0$ is necessary in order for the scalar field to be real. This results in a negative subleading correction to the dS behavior in \cref{eqforu}, contrary to what happens for the Fan $\&$ Wang metric. Another key observation is that  the inequality $\phi'^2 \ll V(x)$ is always satisfied in the dS core, as expected for a gravistar.

Subsequently, one can check for the existence of solutions having a $1/x^4$ term as a subleading correction in $U$, \emph{i.e.}, behaving in the dS core as the Bardeen black hole (see \cref{regularmetricr0}). However, by setting $a_1=a_2=a_3=0$ and $a_4\neq 0$ in \cref{eqp}, one finds, using \cref{Vgeneral}, that reality of both $U$ and $\phi$ rules out this possibility.

Finally, let us now look for solutions having an $\mathcal{O}(x^{-5})$ subleading term in $U$, \emph{i.e.}, behaving as the Hayward black hole. This can be done by setting $a_1=a_2=a_3=a_4=0$ and $a_5\neq 0$ in \cref{eqp}.
Using \cref{solutiongeneral}, one finds the following form of $U$
\begin{equation}\label{Uhayward}
U(x) \simeq 1 - \frac{\hat L^2r_0^2}{x^2} + \frac{14\, a_5}{3r_0\, x^5} + \mathcal{O}(x^{-6})\, ,
\end{equation}
where $\hat L$ is given in terms of $r_0,\, a_5$ and $c_2$ (the explicit expression is rather involved and not particularly illuminating, therefore we do not report it here). The scalar-field kinetic term is, once again, subleading with respect to the potential. By imposing  $a_5<0$ to ensure that the scalar field is real, the desired correction in $U$ has the opposite sign with respect to the one in the Hayward metric (see \cref{regularmetricr0}).

Summarizing, only solutions having Fan $\&$ Wang- and Hayward- like subleading terms to the dS core are allowed. These, however, come out with an opposite sign with respect to the standard black-hole solutions in \cref{regularmetricr0}.   
This result is not accidental and, indeed, we will come back to this point in \cref{sec:NoGodS} while discussing the existence of minima of the function $U$.

\subsection{Asymptotic solution at spatial infinity}
\label{subsec:Aroundinfty}

Let us now discuss the form of the asymptotic solutions at $x\to0$ (spatial infinity). To make contact with asymptotically-flat regular models, we impose that our solution behaves as the Schwarzschild one in this regime. Then we derive the subleading terms allowed by \cref{solutiongeneral}. 

A leading Schwarzschild behavior at spatial infinity requires the following
\begin{equation}\label{asmetric}
P=1, \qquad  U= 1- \frac{2GM x}{r_0}, \qquad V=0 \, 
\end{equation}
to hold at leading order.

Having in mind the well-known regular black-hole solutions mentioned earlier, we restrict our considerations to the first subleading terms in $U$ having the following behavior (again written as functions of the radial coordinate $r$)

\begin{align}
& \text{Fan\& Wang} &U_\text{FW}(r) &= 1-\frac{2GM r^2}{(r+\ell)^3}\simeq 1-\frac{2GM}{r}+\frac{6 G M \ell}{r^2} + \mathcal{O}(r^{-3}),\nonumber\\
& \text{Bardeen} &U_\text{B}(r) &= 1-\frac{2GM r^2}{(r^2 + \ell^2)^{3/2}} \simeq 1-\frac{2GM}{r} + \frac{3GM \ell^2}{r^3} + \mathcal{O}(r^{-4}),\label{assol}\\
& \text{Hayward} &U_\text{H}(r) &= 1-\frac{2GM r^2}{r^3 + \ell^3} \simeq 1-\frac{2GM}{r} + \frac{2GM \ell^3}{r^4} + \mathcal{O}(r^{-5})\, .\nonumber
\end{align}

Similarly to what we did before, we write the most general expansion of $P(x)$, but now near $x=0$, as
\begin{equation}\label{eqpinf}
P(x) = 1 + \frac{b_1}{r_0} x+ \frac{ b_2 }{r_0 } x^2 + \frac{ b_3}{r_0} x^3 + ...
\end{equation}
where the coefficients $b_i$'s have, again, dimensions of length. We will consider the expansion \eqref{eqpinf} only up to terms of order $\mathcal{O}(x^3)$, which are enough to generate terms up to order $\mathcal{O}(x^4)$ in the expansion of $U$.

\Cref{eqpinf}, together with \cref{solutiongeneral}, gives a simple way to characterize the solutions endowed with a nontrivial scalar field, \emph{i.e.}, with a scalar charge \footnote{More precisely, we define the scalar charge as the coefficient of the leading $1/r^n \sim x^n,\, n=1,2 ...$ term in the $r\to \infty$ expansion of the scalar field $\phi$.}.   
First, \cref{phigeneral} tells us that the leading term in the expansion of $\phi$ is set by $b_2$, which therefore determines its $x$ charge: $\phi\sim \sqrt {-b_2} \, x$. Second, \cref{Ugeneral} implies that the coefficient of the subleading term in $U(x)$ is proportional to $b_1 GM x^2$. Moreover, the leading term in the potential is set by $b_3$ and scales as $V\sim b_3 x^5$. Therefore, the asymptotic behavior of the potential at $x\to 0$ 
\begin{equation}    
V\sim x^5\sim \phi^5
\end{equation}
is universal for static, asymptotically-flat, spherically-symmetric solutions endowed with a scalar field which decays as $x$ for $x\to 0$. Notice that this result is independent of the behavior of the solution in the core and, in fact, it has been observed also for asymptotically-flat, singular black-hole solutions \cite{Cadoni:2015gfa}. 
In particular, this universality is implied by the presence of a kinetic term that dominates asymptotically over the potential. This regime is therefore well described by a conformally-invariant source.

We have now to distinguish between the two cases: $b_1\neq 0$ and $b_1=0$. In the former case, we get solutions whose metric part behaves asymptotically as the Fan $\&$ Wang metric, with a subleading term of order $x^2$ both in $\phi$ and $U$ and a quintic potential $V\sim x^5\sim \phi^5$. More specifically, choosing $b_3=0$ for simplicitly, we get for $U$ and $\phi$
\begin{equation}
U(x) \simeq 1-\frac{2GM}{r_0}x + \frac{2b_1GM}{r_0^2}x^2 +\mathcal{O}(x^3), \quad \phi(x) \simeq -\frac{\sqrt{-2b_2}}{\sqrt{r_0}} \, x+ \frac{b_1 \, \sqrt{-b_2}}{2\sqrt{2}\,  r_0^{3/2}}\, x^2+\mathcal{O}(x^3)\, ,
\end{equation}
whereas the potential reads as 
\begin{equation}
V(x) \simeq \frac{4b_2 (-3b_1+4GM)}{5r_0^4}\, x^5 +\mathcal{O}(x^7)\sim  \phi^5\, .
\end{equation}
As a side remark, we note that these results shed light on the discussion of Ref. \cite{Cadoni:2023nrm} about the conformal symmetries of the Fan $\&$ Wang solution. The parameters $2GM$ and $b_2$ determine the two charges (gravitational and scalar) that appear as harmonic functions in the Poisson equation $\nabla^2 \Phi= 4 \pi \rho$. The $b_1$ parameter, instead, determines the charge associated with the $1/r^2$ term producing the (conformal) density $\rho \propto 1/r^4$ \cite{Cadoni:2023nrm}. We emphasize that this same parameter appears as a ``quantum hair'', $\ell$, in the nonsingular black-hole metric, which here acquires a geometric interpretation in terms of the constant mode in the radial function $R(r)=r+ b_1$.

When, instead, the coefficient $b_1$ vanishes, the subleading quadratic terms in $U(x)$ and $\phi(x)$ are not present anymore. The leading linear term in $\phi$ is now always determined by $b_2$. Depending on the value of the parameter $b_3$, we will have  subleading $\mathcal{O}(x^3)$ or $\mathcal{O}(x^4)$ terms. Specifically, for $b_1=b_3=0$ and $b_2\neq 0$ we get 
\begin{equation}\label{Ucase3}
U(x) \simeq 1-\frac{2GM}{r_0}\, x + \frac{4 b_2 G M}{5r_0^2} \, x^3 + \mathcal{O}(x^4), \qquad \phi(x) \simeq \sqrt{-\frac{2b_2}{r_0}}\, x -\frac{b_2\sqrt{-\frac{b_2}{r_0}}}{3\sqrt{2} r_0}\, x^3 + \mathcal{O}(x^4)\, ,\, 
\end{equation}
whereas the potential is
\begin{equation}
V(x) \simeq \frac{16 b_2 GM}{5 r_0^4} \, x^5 - \frac{4 b_2^2}{r_0^4}\, x^6 + \mathcal{O}(x^7)\, .
\end{equation} 
The first subleading correction in $U$ is of order $\mathcal{O}(x^3)$, just like the Bardeen metric.
Again, in order to have a real scalar field, $b_2<0$ so that the $\mathcal{O}(x^3)$-order correction in \cref{Ucase3} is negative, contrary to what happens for the Bardeen metric. Notice that there could be contributions to the $\mathcal{O}(x^3)$ term in $U$ coming from higher order terms in $P(x)$, potentially correcting the sign of this term and making it positive. On the other hand, if $b_1=0$, $b_3\neq 0$, and $b_2\neq 0$, the coefficient of the $\mathcal{O}(x^3)$ term in $U$ would get a contribution from $b_3$. A brief calculation shows that this would not be able to make the sign of the $\mathcal{O}(x^3)$ correction in $U$ positive as desired.
Finally, choosing the particular value $b_3=-2 b_2GM$, the coefficient of the $\mathcal{O}(x^3)$ term in $U$ vanishes, leaving an $\mathcal{O}(x^4)$ subleading term in $U$,
\begin{equation}
U(x) \simeq 1-\frac{2GM}{r_0}\, x + \frac{b_2^2}{3r_0^2}\, x^4 + \mathcal{O}(x^5)\, .
\end{equation}
The subleading correction $\mathcal{O}(x^4)$ is surely positive and has the same asymptotic behavior of the Hayward black hole. 

One can easily check that all the asymptotic $x\to 0$ solutions discussed in this subsection describe a regime in which the kinetic term of the scalar field dominates over the potential, \emph{i.e.}, $V\ll \phi'^2$. This is a consequence of the asymptotic behavior $\phi\sim x$ and $V\sim  x^5$, again, for $x\to 0$.

\section{Nonexistence theorem for smooth interpolating solutions}
\label{sec:NoGodS}

In the previous section, we investigated the asymptotic behavior of the solutions of Einstein-scalar gravity,  separately, in the dS core and in the asymptotically-flat region. The next step is to look for smooth solutions of the theory interpolating between the two $r=0$ and $r=\infty$ asymptotics. A first indication that the existence of smooth interpolating solutions is a rather involved issue for minimally-coupled scalar fields comes from the ``wrong'' signs of the subleading terms found in \cref{subsec:Aroundzero}.  

We will tackle the problem of the existence of smooth interpolating solutions in full generality by resorting to the dynamical equations \eqref{solutiongeneral}. 
We will prove a general theorem stating that the theory \eqref{action} does not admit smooth solutions interpolating between a dS spacetime near the core and Schwarzschild at asymptotic infinity. Our theorem can be considered as an extension of the no-go theorems, first proved by Bronnikov and Shikin \cite{Bronnikov:2001ah,Bronnikov:2001tv} and Gal'tsov and Lemos \cite{Galtsov:2000wmw} concerning asymptotically-flat smooth solutions endowed with horizons, to spacetimes without event horizons. 
In the following, thus, we will restrict our considerations to the latter. 

Let us consider the metric function $U(r)$. As stated above, we require the object's core to be described by the dS spacetime, and to be asymptotically flat at infinity. As a result, close to $r\to0$, the metric function starts at $U(0)=1$ and then decreases as the radial coordinate grows. Moreover, the metric function must increase asymptotically with $r$ since $U(r\to\infty)\to 1$ from below. Smoothness of $U$, then, implies the presence of (at least) a local minimum in the region $0<r<\infty$. Owing to the monotonicity of the reparametrization function and of its derivative (the change in the radial coordinate from $r$ to $x$), also $U(x)$ must have a minimum for $0<x<\infty$.   
We will now show that \cref{solutiongeneral} forbids the presence of such a minimum, thus preventing the existence of smooth interpolating solutions between the two abovementioned asymptotic behaviors. Throughout the proof, we will exclude the presence of singularities for $U(r)$ and, therefore, $U(x)$.

We start from \cref{Ugeneral}, we compute its first derivative with respect to $x$ and assume that $U(x)$ has an extremum $U_0=U(x_0)\neq 0$ for some $x_0$ in the region $0<x_0<\infty$ (we are, thus, excluding $x_0=0,\infty$), where $\dd U/ \dd x = 0$. Evaluating the resulting equation at $x=x_0$ we get (dot will refer to derivation with respect to $x$) 
\begin{equation}\label{U0}
U_0 = \frac{2+\frac{c_1}{r_0}x_0}{2 P_0^2 -2 x_0 P_0 \dot P_0}\, ,
\end{equation}
where the subscript ${}_0$ refers to quantities calculated at $x = x_0$. This convention will be adopted in the reminder of this section. Notice that $P(x)$ must be monotonic and strictly positive in order to avoid spacetime singularities in \cref{Ugeneral}, thus $P_0> 0$ and $\dot P_0>0$ hold.

As we are excluding the presence of horizons, $U$ cannot change its sign anywhere. Additionally, if we restrict to asymptotically-flat solutions, it must be positive everywhere. Therefore
\begin{equation}\label{bound}
\frac{2+\frac{c_1}{r_0}x_0}{2 P_0^2 -2 x_0 P_0 \dot P_0} >0\, .
\end{equation}

As we discussed in \cref{GRCoupledScalarFieldExactSolution}, the integration constant $c_{1}$ is always related to the asymptotic behavior at infinity, namely it is necessarily proportional to minus the ADM mass. Therefore, the numerator in \cref{bound} is equal to  the Schwarzschild metric function increased by one. By imposing an asymptotic Schwarzschild behavior at spatial infinity for $U(x)$, together with its positivity in the whole range $0<x<\infty$, the numerator in \cref{bound} must be positive, which in turn implies the positivity of the denominator by virtue of \cref{bound}. 

In order to assess the nature of the extremum $x_0$ (either maximum or minimum), we evaluate the second derivative of $U$ at $x_0$. We obtain
\begin{equation}
\begin{split}\label{ddotU0}
&\ddot U_0 P_0^2 x_0 = \frac{c_1}{r_0} -2U_0 P_0 \dot P_0 + 2U_0 \dot P_0^2 x_0 + 2 U_0 P_0 \ddot P_0 x_0 \, .
\end{split}
\end{equation}

$P_0^2 x_0$ is surely positive and will not alter the sign of $\ddot U_0$ on the left hand side. Moreover, $c_1$ is either zero or negative (proportional to $-M$), while $\ddot P_0 < 0$ to ensure the scalar field to be real (see \cref{phigeneral}). To analyze the remaining two terms, we note that we can write them as
\begin{equation}
-2U_0 P_0 \dot P_0 + 2U_0 \dot P_0^2 x_0 = -\frac{U_0 \dot P_0}{P_0}\left(2 P_0^2 - 2x_0 P_0 \dot P_0 \right)\, .
\end{equation} 
The quantity in round brackets exactly corresponds to the denominator of \cref{U0}, which, according to the discussion below \cref{bound}, must be positive. Since also $U_0 >0$, the quantity on the right hand side is negative ($P$ is monotonically increasing as $x$ grows, \emph{i.e.}, $\dot P >0$ everywhere), and so is the quantity on the left hand side. From this reasoning, it follows that $\ddot U_0<0$, implying that if it exists a point $x = x_0$ such that $\dot U_0 = 0$, this cannot be a minimum, but only a maximum.

The previous theorem can be easily extended to forbid the existence of smooth solutions interpolating between a dS core and an AdS$_4$ (or even AdS$_2\times$ $S^2$) spacetime at $r\to \infty$. In fact, the asymptotic AdS$_4$ behavior requires, at leading order, $U=1+ r^2/L^2$ (or $U=r^2/L^2+\text{const}$ for AdS$_2$), implying the presence of a minimum for $U$ at finite values of $x$ and $c_1=0$. The same argument used above, then, rules out dS$_4 \rightarrow$ AdS$_4$ (or  $\text{AdS}_2 \times \text{S}^2$) interpolating solutions. On the other hand, our theorem does not forbid smooth solutions which have an AdS instead of a dS core\footnote{A particular example of such solutions was already investigated in Ref.~\cite{Franzin:2018pwz}.} (see \cref{sec:smoothAdScore,sec:ScalarLump} for explicit examples).

Even if our theorem forbids the existence of smooth solutions interpolating between a dS core and the Schwarzschild spacetime at infinity, we can still construct jointed solutions with these asymptotics as we shall show in \cref{sec:JointedSol}. 

\section{Exact solutions with AdS core  and flat asymptotics}
\label{sec:smoothAdScore}

\begin{figure}[!h]
\centering
\subfigure[]{\includegraphics[width= 8 cm, height = 8 cm,keepaspectratio]{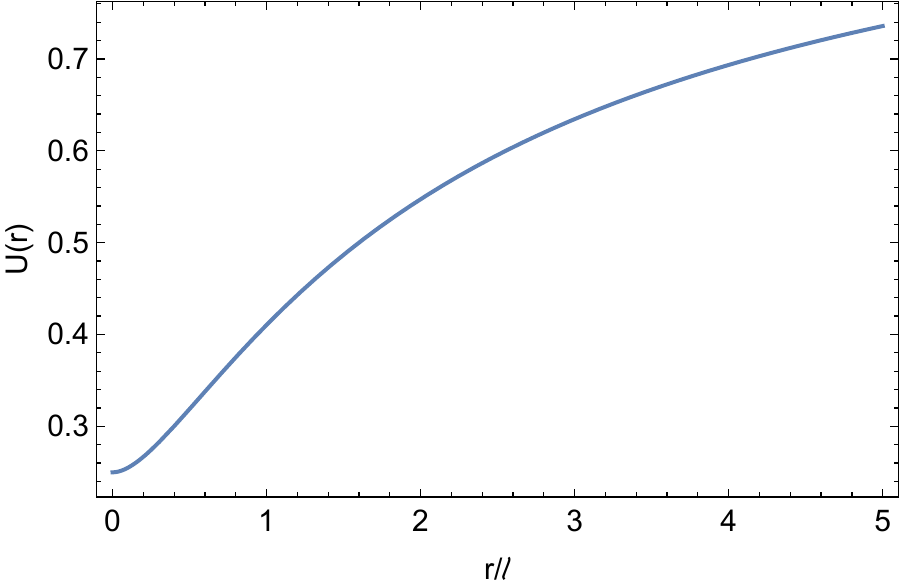} \label{fig:Ufunction}}
\hfill
\subfigure[]{\includegraphics[width= 7.9 cm, height = 7.9 cm,keepaspectratio]{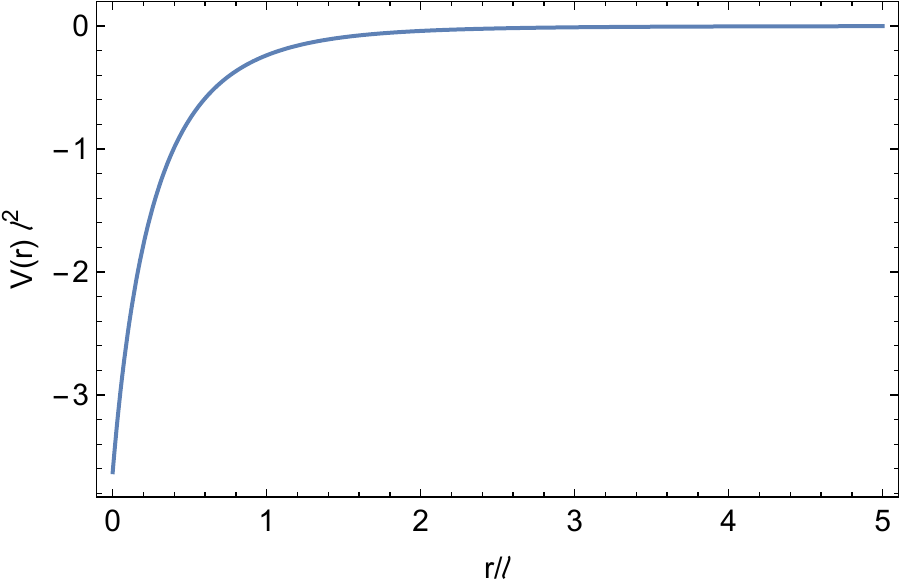} \label{fig:Vfunction}}
\caption{ Metric function $U$ and potential $V$ for  exact solutions with AdS core. {\textbf{Figure (a):}} Metric solution as a function of $r/\ell$. {\textbf{Figure (b):}} Potential as a function of $r/\ell$. }
\end{figure}

In this section we will construct an explicit exact solution interpolating between an AdS spacetime in the core and the Schwarzschild spacetime at $r\to \infty$. This will also allow us to check the validity of the nonexistence theorem proved in the previous section and to explicitly construct a regular solution possessing a $1/r^2$ correction at infinity, mimicking the correction of the Fan $\&$ Wang metric (see \cref{assol}). 

The simplest example allowing for such interpolation is generated by the following ansatz for $P$
\begin{equation}
P(x) = 2-\frac{r_0^3}{(r_0 + \ell \, x)^3} \simeq \begin{cases} & 2-\frac{r_0^3}{\ell^3 x^3}\,, \qquad \qquad \quad \, \, \text{for} \quad x \to \infty \\
&1+\frac{3\ell}{r_0}x - \frac{6 \ell^2}{r_0^2}x^2\,, \qquad \text{for} \quad x \to 0  \end{cases}\, ,
\end{equation}
where $\ell >0$ is a free additional length scale. The radius of the 2-sphere reads
\begin{equation}
R(x) = \frac{r_0}{x}\left[2-\frac{r_0^3}{(r_0+\ell x)^3}\right]\, .
\end{equation}
$R(x)$ reduces to $2r_0/x$ near $x \to \infty$ (so that GR solutions are recovered by a simple rescaling of the coordinates), while at $x \to 0$, $R(x) \to r_0/x$. In the limit $\ell \to 0$, we have $R(x) = r_0/x$ everywhere. $\ell$, therefore, keeps track of the deviations from GR solutions. Notice that, with this form of $P$, the scalar field is real everywhere. 

One can now compute the metric function $U$, and the potential $V$ using  \cref{Ugeneral}. A regular solution requires $c_1 = 0$, while $c_2$ is determined by imposing asymptotic flatness. This last requirement constrains $c_2$ to be positive, giving thus an AdS spacetime in the core. The explicit expressions of $U$, $V$ and $c_2$ are cumbersome, so that we do not write them here. We just show the plots for $U$ and $V$ in \cref{fig:Ufunction} and \cref{fig:Vfunction} respectively, as a function of the radial coordinate $r$. In particular, the potential reduces to a negative constant near $r\sim 0$ at leading order, consistently with the AdS behavior. 

Let us conclude this section with some comments. Firstly, a consequence of having set $c_1 = 0$ is the absence of an event horizon, so this solution represents a regular horizonless compact object. Secondly, the requirement of asymptotic flatness requires $c_2$ to be positive, \emph{i.e.},  the core of the object can never be given by a dS vacuum: this represents a particular check of our nonexistence theorem proved in \cref{sec:NoGodS}.

\section{Jointed solutions}
\label{sec:JointedSol}

The nonexistence theorem proved in \cref{sec:NoGodS} rested on the smoothness of the interpolating  solution between two vacua: the dS and the Minkowski ones. We can, however, circumvent the conclusions of the theorem by constructing jointed solutions through the Israel-Lanczos conditions \cite{Israel:1966rt}.

We will consider two different solutions of the field equations, in the interior and exterior of a hypersurface $\Sigma$ onto which the junction is performed. In particular, inspired by the role of the scale symmetry in some relevant regular black-hole models \cite{Cadoni:2023nrm} (see the discussion at the end of \cref{subsec:inflstars}), we will consider Einstein's gravity sourced by a conformal field theory for the exterior solution, namely a scalar-field theory with an identically zero potential. 

\subsection{Interior solution}
\label{subsec:interiorsolution}

In the interior, we look for a solution behaving as the dS spacetime near $r \sim 0$. The computations of \cref{subsec:Aroundzero} imply that, near $x \to \infty$, the subleading correction to $1$ in the function $P$ (the contribution giving the dS solution) must be at least of order $\mathcal{O}(x^{-3})$. 

There are actually many solutions characterized by this core behavior. We will select a simple, illustrative example which allows for an exact solution. A simple choice of the radial metric function allowing to recover the abovementioned behavior in the core is the following
\begin{equation}
R(x) = \frac{r_0/x}{\sqrt{1+\left(\frac{r_0}{\ell x}\right)^\alpha}}\, ,
\end{equation}
where $\ell$ is, again, a length-scale parameter, while $\alpha$ must be at least equal to $3$ to ensure a dS-like behavior in the core. For simplicity reasons, we will here select $\alpha = 4$. Therefore
\begin{equation}
P(x) = \frac{\ell^2 x^2}{\sqrt{r_0^4 + \ell^4 x^4}}\, .
\end{equation}
At $x \to \infty$, $P(x) \simeq 1 - r_0^4/2 \ell^4 x^{4} + \mathcal{O}(x^{-5})$. 
Concerning the metric function, $c_1$ must be fixed equal to zero for the solution to be regular, as done in \cref{sec:smoothAdScore}. $U$ reads as 
\begin{equation}\label{Uinner}
U(x) = \frac{3\ell^8 x^8 + 3c_2 \ell^8 r_0^2 x^6-6 \ell^4 r_0^4 x^4 - r_0^8}{3\, \ell^4 \, x^4 (r_0^4 + \ell^4 \, x^4)}\, .
\end{equation}
At $x \to \infty$, it behaves as $U(x) \simeq 1+ c_2 r_0^2/x^{2}+\mathcal{O}(x^{-4})$, so $c_2 < 0$ in order to have a dS spacetime. 
The scalar field reads as
\begin{equation}
\phi(x) = 2\sqrt{3} \tan^{-1}\left(\frac{\sqrt{-r_0^4 + 5 \ell^4 x^4}}{\sqrt{6} \, r_0^2} \right) -\sqrt{2}\tan^{-1} \left(\frac{\sqrt{-r_0^4 + 5 \ell^4 x^4}}{r_0^2}  \right)\, .
\end{equation}
At $x \to \infty$, we have $\phi(x) \simeq \phi_0 -\sqrt{10} r_0^2/(\ell^2 \, x^2) + \mathcal{O}(x^{-3})$, with $\phi_0 \equiv \frac{2\sqrt{3}-\sqrt{2}}{2}\pi$. 

Finally, the potential is
\begin{equation}
V(x) = -\frac{2 \left(3 x^4 \ell ^4-r_0^4\right) \left[3 c_2 x^6 \ell ^8 \left(x^4 \ell ^4-5 r_0^4\right)+2 r_0^2 \left(10 r_0^4 x^4 \ell ^4+r_0^8-15 x^8 \ell ^8\right)\right]}{3 x^2 \ell ^4 \left(r_0^4+x^4 \ell ^4\right)^3}\, .
\end{equation}
At $x \to \infty$, we have $V(x) \simeq -6c_2 + 60 r_0^2/(\ell^4 \, x^2) + \mathcal{O}(x^{-3})$.

As expected, if $c_2 <0$, the leading term is a positive constant value (equal to a positive cosmological constant) in the core, consistent with a dS solution.

Although we assume the present solution is not valid in the regime $x\to0$, we note that, in this limit, it has a quite singular behavior, since the radius of the two sphere goes to zero, whereas both $U$ and $V$ diverge as $1/x^4$ and as $1/x^2$ respectively. 
 
\subsection{Exterior solution}
\label{subsec:exteriorsolution}

In the exterior, we consider the simple case $V = 0$, for which the theory is conformal. 
Exact, asymptotically-flat solutions of Einstein's gravity coupled to a scalar field with $V=0$ are already know \cite{Cadoni:2015gfa}. Using the solution generating algorithm \eqref{solutiongeneral}, they are generated by considering $P(x) = (1-x)^\beta$, where $\beta$ is a dimensionless parameter constrained by $0< \beta < 1$, in order for the scalar field to be real and nonzero. The full solution reads, restricting to the exterior region $x<1$, \emph{i.e.}, $r > r_0$ (see below),

\begin{equation}\label{mas}
R(x) = \frac{r_0}{x}(1-x)^\beta,\qquad \phi(x) = -2\sqrt{\beta\left(1-\beta \right)}\, \ln\left(1-x \right), \qquad 
U(x) = (1-x)^{1-2\beta}\, .
\end{equation}

The expansion in the asymptotic, $x\to 0$ region reads as
\begin{equation}\label{ae}
\phi(x) \simeq 2\sqrt{\beta(1-\beta)} \, x + \mathcal{O}(x^2), \qquad U(x) \simeq 1+(2 \beta -1) x +\mathcal{O}(x^2), \qquad  R(x)\simeq
\frac{r_0}{x}.
\end{equation} 
We immediately see that this solution has an event horizon at $x = 1$, namely at $r = r_0$, which is related to the classical Schwarzschild radius (or equivalently to the ADM mass). Indeed, we can immediately read the latter from the linear term in the expansion for $U$ given by \cref{ae}
\begin{equation}\label{RSexternal}
R_\text{S} = (1-2\beta) \, r_0\, .
\end{equation}
The positivity of $R_\text{S}$ further constrains $\beta$ to the interval $0<\beta <\frac{1}{2}$, where $\beta=1/2$ is excluded since this would give a trivial (constant) $U$ according to \cref{mas}.
As we shall see in the following subsections, the presence of the event horizon at $x=1$ is irrelevant for our purposes, since the junction point between the interior and exterior solutions will be outside this horizon, namely in the $x<1$ region. 

\subsection{Continuity conditions of the induced metric}
\label{subsec:continuitycondition}

Let us now joint the core solution, considered in \cref{subsec:interiorsolution}, with the exterior solution of \cref{subsec:exteriorsolution} using the Israel formalism \cite{Israel:1966rt}. This can be done by imposing the continuity of the induced metric $h_{ij}$ onto the $3$-dimensional hypersurface $\Sigma$, located at the radial position $r = y_0$, at which the two solutions are jointed \footnote{In the following, latin indices will be used to refer to quantities projected onto the hypersurface.}. 

In general, the conditions for a smooth joining of two metrics are 
\begin{equation}
[h_{ij}] =0\,,  \qquad [K_{ij}] = 0\, ,
\end{equation}
where $K_{ij}$ is the extrinsic curvature and  $[...]$ indicates the discontinuity of the quantities across the hypersurface $\Sigma$, namely\footnote{We shall indicate quantities in the exterior and in the interior of $\Sigma$ with the superscripts ``$+$'' and ``$-$'' respectively.} $[A] \equiv A(y_0^+) - A(y_0^-)$.

However, only the left one is necessary. If the extrinsic curvature is not continuous across $\Sigma$, a thin shell with nonzero surface stress-energy tensor is generated, which is described by the so-called Lanczos equations
\begin{equation}\label{LanczosSET}
S_{ij} = -\frac{\epsilon}{8\pi}\left([K_{ij}]-[K]h_{ij} \right)\, ,
\end{equation}
where $\epsilon = n_\mu n^\mu$, with $n_{\mu}$ the unit normal to $\Sigma$ pointing from the interior to the exterior, according to Israel's prescription (more details can be found, \emph{e.g.}, in Refs.~\cite{Visser:2003ge,Poisson:2009pwt}).

In this section, we will discuss the first condition $[h_{\mu\nu}]=0$. By projecting the $4$-dimensional metric, $\dd s^2 =-U \dd t^2 + U^{-1} \, \dd r^2 + R^2 \dd \Omega^2 $, onto the hypersurface at constant $r$, namely $r = y_0$, we obtain the induced metric 
\begin{equation}
h_{\mu\nu} = \text{diag} \left(-U(y_0), \, R(y_0)^2, \, R(y_0)^2 \sin^2 \theta\right)\, .
\label{Inducedmetric}
\end{equation}
The following step is to analyze the conditions and the values of the parameters $\ell$, $r_0$ and $c_2$ necessary to have a well-defined and admissible $y_0$. The continuity of the induced metric will provide us with only two equations for the two metric functions, $U$ and $R$, and four unknowns, $\ell$, $r_0$, $y_0$ and $c_2$. Notice that $\beta$ is not considered among the latter since it should be fixed a priori. For convenience, we express $\ell$ and $c_2$ in terms of $r_0$ and $y_0$. Using the continuity condition for $R$, $R^+=R^-$ one can easily get $\ell$:
\begin{equation}
\ell = \frac{y_0\, \gamma^{1/4}}{\left(1-\gamma \right)^{1/4}}\, , \qquad \, \gamma \equiv \left(1-\frac{r_0}{y_0} \right)^{2\beta}\, .
\label{ljointed}
\end{equation}
To ensure the positivity of $\ell$, we require $0 < \gamma < 1$, which translates to $y_0 > r_0$. In this way, the shell now replaces the horizon of the exterior metric, as it is usually the case for jointed solutions leading to horizonless compact objects, such as gravastars. 

We can now compute $c_2$ as a function of $r_0$ and $y_0$, by using the continuity condition for the metric function $U$, $U^+=U^-$. Employing also \cref{ljointed}, we get 
\begin{equation}\label{c2jointed}
c_2 = \frac{\left(1-\frac{r_0}{y_0}\right)^{-4 \beta } \left[4 \left(\left(1-\frac{r_0}{y_0}\right)^{2 \beta }-2 \left(1-\frac{r_0}{y_0}\right)^{4 \beta }+1\right)-3 \frac{r_0}{y_0}\right]}{3 y_0^2}\, .
\end{equation}

As seen in \cref{subsec:interiorsolution}, $c_2<0$ to have a dS spacetime near $r \sim 0$. The quantities outside the square brackets are surely positive, so the sign of $c_2$ is determined by the sign of the function
\begin{equation}\label{c2expressionbeta}
\mathcal{F}\left(\frac{r_0}{y_0}\right)\equiv 4 \left(1-\frac{r_0}{y_0}\right)^{2 \beta }-8 \left(1-\frac{r_0}{y_0}\right)^{4 \beta }+4 -3 \frac{r_0}{y_0}\, .
\end{equation}

\begin{figure}[h!]
\centering
\includegraphics[width= 9 cm, height = 9 cm,keepaspectratio]{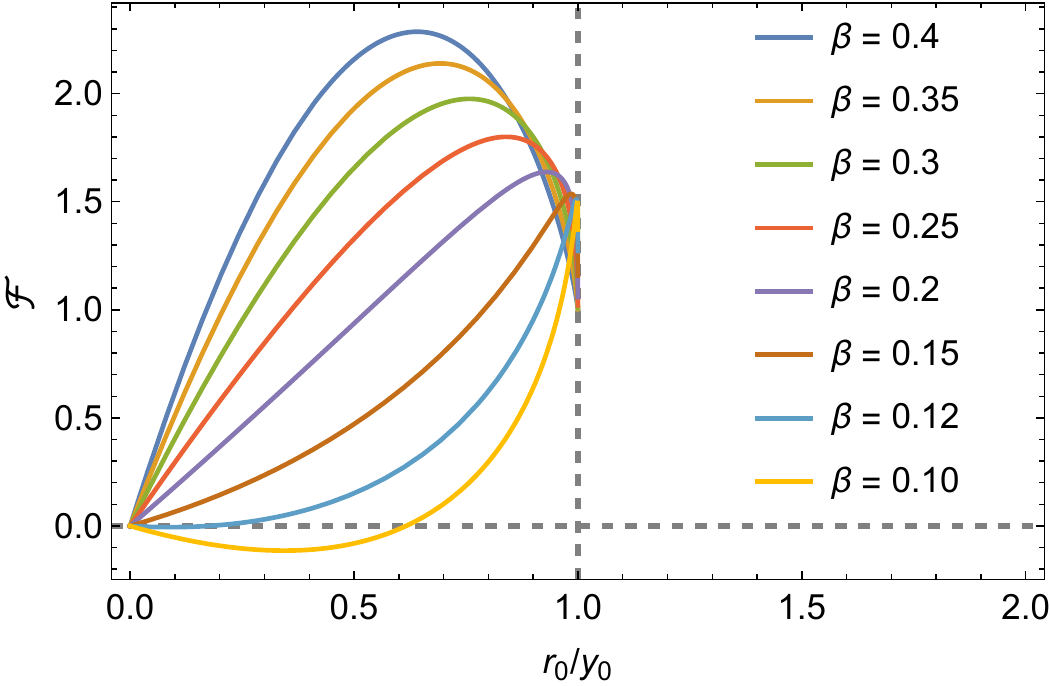}
\caption{Plot of the function $\mathcal{F}(r_0/y_0)$ \eqref{c2expressionbeta} for different values of $\beta$. Since $r_0 < y_0$, the plot stops at $1$. We note that only for values of $\beta \leq 0.1$, $c_2$, given by \cref{c2jointed}, is negative. }
	\label{fig:c2expressionbeta}
\end{figure}

\begin{figure}[!h]
\centering
\subfigure[]{\includegraphics[width= 8 cm, height = 8 cm,keepaspectratio]{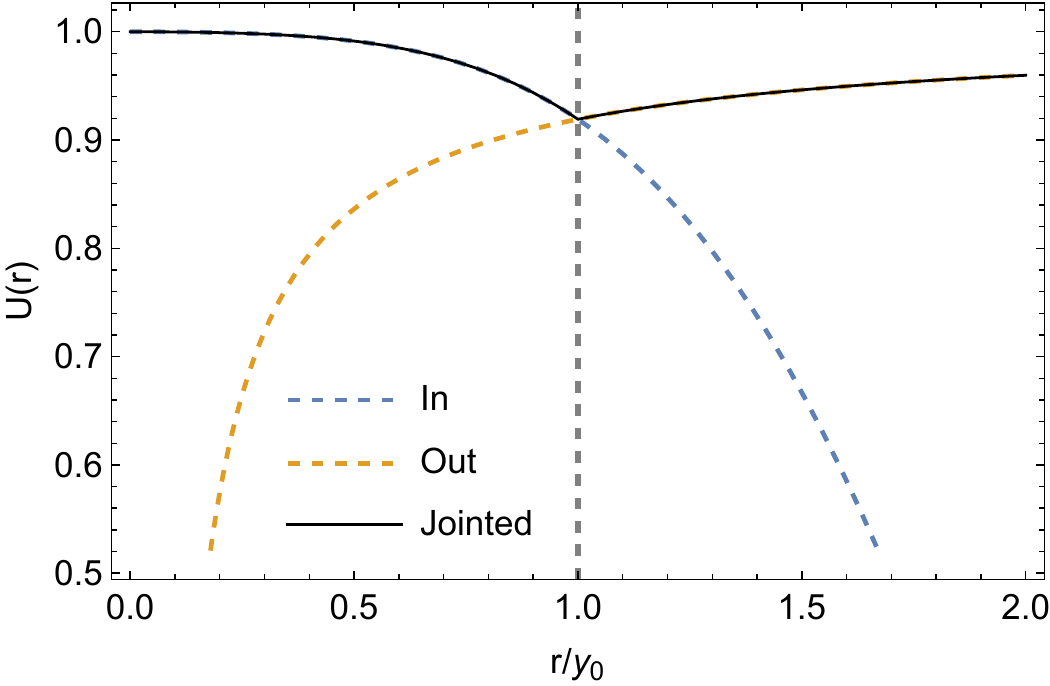} \label{fig:fjointed}}
\hfill
\subfigure[]{\includegraphics[width= 7.9 cm, height = 7.9 cm,keepaspectratio]{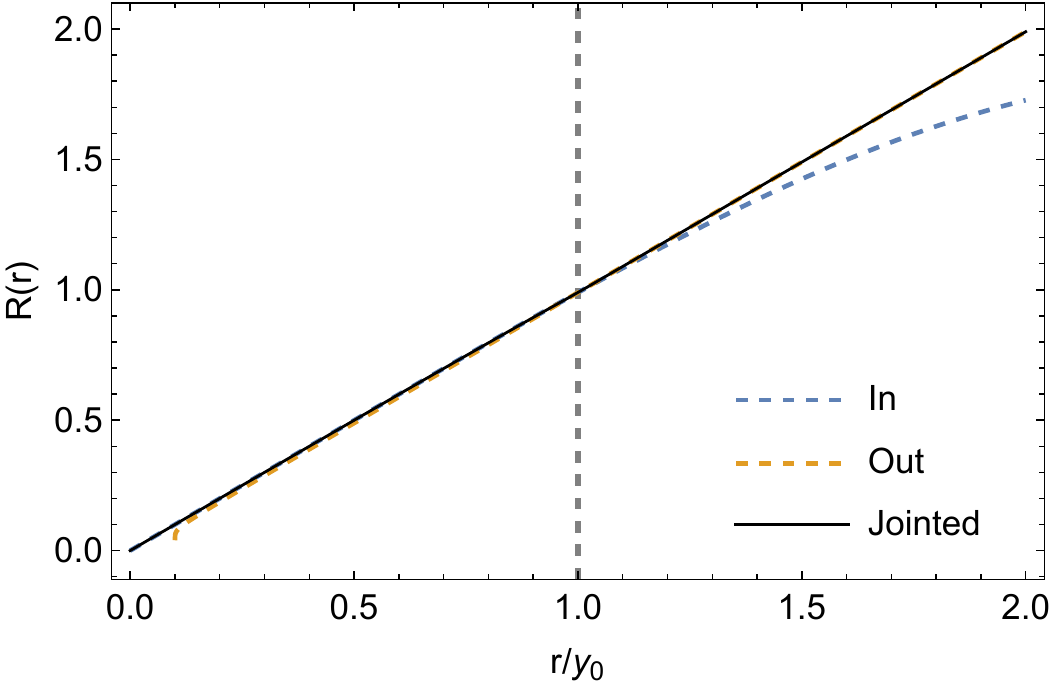} \label{fig:Rjointed}}
\caption{{\textbf{Figure (a):}} junction of the two metric functions  $U$ for selected values of the parameters. {\textbf{Figure (b):}} junction of the two radial function for selected values of the parameters.
For both figures, the inner solution is the dashed blue line, while the exterior one is given by the dashed orange line. The black solid line corresponds to the full jointed solution. The vertical line, instead, corresponds to the position $y_0$ of the shell. We set $\beta = 0.1$, $r_0/y_0 = 0.1$. }
\end{figure}

The parameter region of interest is determined by the inequalities $\mathcal{F}\left(r_0/y_0\right)<0$ and $r_0 / y_0 < 1$. In \cref{fig:c2expressionbeta} we plot the function $\mathcal{F}(r_0/y_0)$, from which one can infer the allowed parameter space. It is in this region that the interior solution, characterized by the dS vacuum near the core, can be jointed with continuity to the exterior solution, characterized by the Minkowski vacuum at infinity. We plot an explicit example for selected values of the parameters in \cref{fig:fjointed,fig:Rjointed}. Finally, we stress that, in general, the potential and the scalar field will be discontinuous at the junction.

\subsection{Geometric quantities and the stress-energy tensor of the shell}
Let us now characterize the stress-energy tensor of the shell using the Lanczos equations \eqref{LanczosSET}. 
First of all, the normal to the hypersurface $\Sigma$ reads, for static, spherically-symmetric spacetimes
\begin{equation}
n_\mu = \frac{\delta^r_{\mu}}{\sqrt{g^{rr}}}\, ,
\end{equation}
on both sides of the shell. We also fixed $n^r >0$ to have the normal directed from the interior to the exterior (according to Israel's prescription) and, thus, $n_\mu n^\mu = \epsilon = 1$. It is easy to verify that the definition of the induced metric $h_{\mu\nu} = g_{\mu\nu}-n_\mu n_\nu$, with these choices, is consistent with \cref{Inducedmetric}. Notice that the normal vector changes while crossing the shell radially since $g^{rr}$ changes from the inside to outside the object. For this reason, we define $n^{\pm}_{\mu}$ as the normal relative to the components outside (plus sign) and outside (minus sign) of the shell.

\begin{figure}[!h]
\centering
\subfigure[]{\includegraphics[width= 8 cm, height = 8 cm,keepaspectratio]{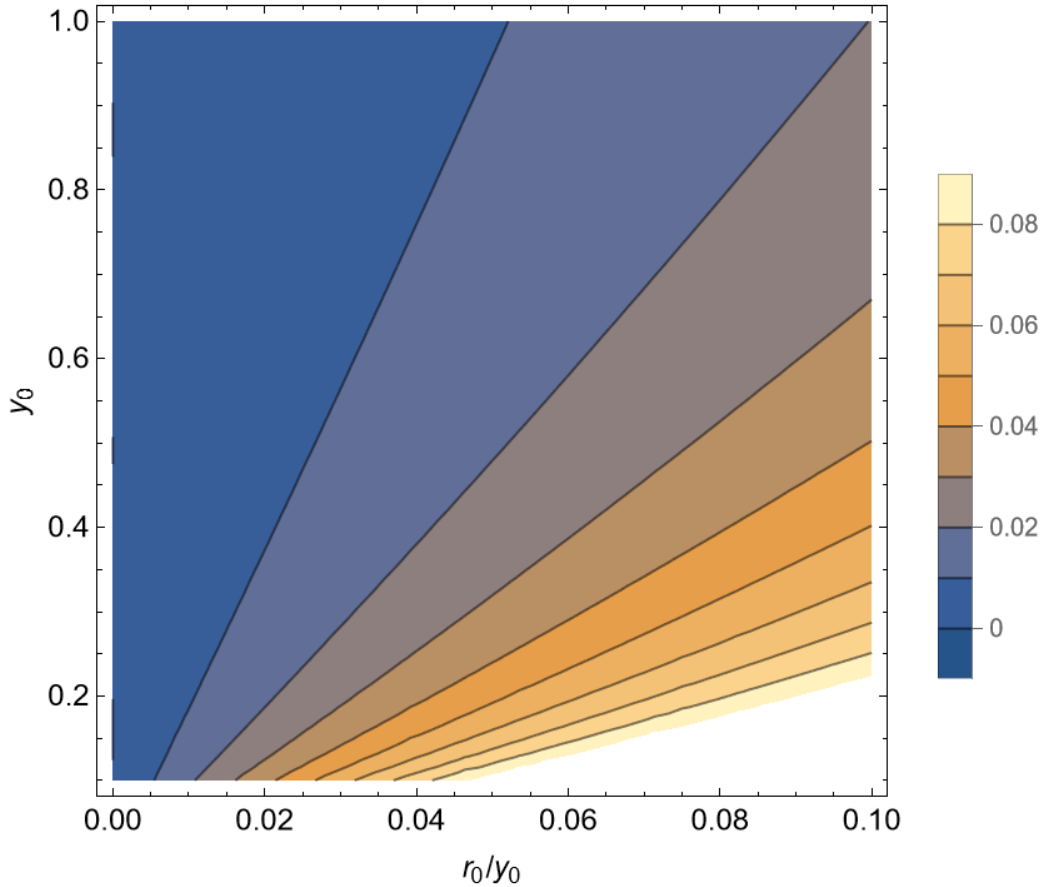} \label{fig:3DWEC}}
\hfill
\subfigure[]{\includegraphics[width= 7.9 cm, height = 7.9 cm,keepaspectratio]{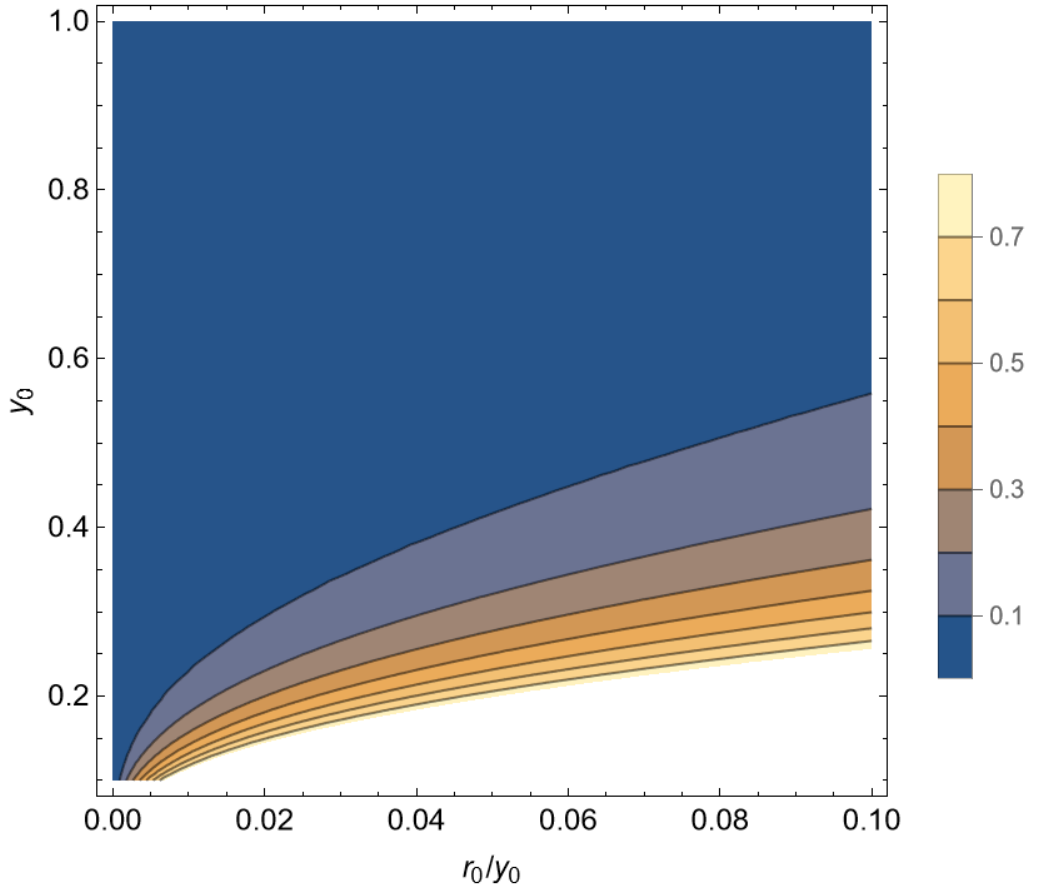} \label{fig:3DNEC}}
\caption{Contour plot for surface energy density $\sigma$ and pressure  $\mathcal{P}$ of the shell. \textbf{Figure (a):} Contour plot for the surface energy density $\sigma(y_0)$ as a function of the parameters $y_0$ and $r_0$. \textbf{Figure (b):} Contour plot for  $\sigma(y_0) + \mathcal{P}(y_0)$ as a function of the parameters $y_0$ and $r_0$.\\
For both figures, $\beta = 0.1$, while the others are fixed to require continuity of the induced metric (see \cref{subsec:continuitycondition}). As it can be seen, both the WEC and NEC are satisfied in these intervals of parameters. }
\end{figure}

We then compute the components of the extrinsic curvature, whose general expression is given by
\begin{equation}\label{Extrinsiccurvaturegeneral}
K_{\mu\nu} = \frac{1}{2}\nabla_\mu n_\nu + \frac{1}{2}\nabla_\nu n_\mu\, = \frac{1}{2}\partial_\mu n_\nu + \frac{1}{2}\partial_\nu n_\mu-\Gamma^\lambda_{\mu\nu} n_\lambda\, .
\end{equation}
Also in this case, $K_{\mu\nu}$ has different components in the interior and exterior, which will determine the magnitude of the discontinuity of the extrinsic curvature across the shell $[K_{ij}]$. The calculations and the explicit expressions of the extrinsic curvature components and their traces, in the interior and exterior, evaluated at $r = y_0$, are reported in \cref{AppA:ExtrinsicCurvature}.

With these results, we can compute the components of the stress-energy tensor of the shell \eqref{LanczosSET}. Beyond their geometrical meaning, these components have also a physical interpretation in terms of the properties of the shell, once \cref{LanczosSET} is interpreted as the stress-energy tensor of a perfect fluid
\begin{equation}\label{SETshellcomponents}
S^i_j = -\frac{1}{8\pi}\left([K^i_j] - \delta^i_j [K] \right) = \text{diag}\left(-\sigma, \mathcal{P}, \mathcal{P} \right)\, ,
\end{equation}
where $\sigma$ is the surface energy density, while $\mathcal{P}$ is the surface pressure of the shell. Their explicit expressions, evaluated at $r = y_0$, are reported in \cref{AppA:ExtrinsicCurvature} as well.

We finally check if the surface energy and pressure satisfy the usual energy conditions of thin shells \cite{Visser:2003ge}, \emph{i.e.}, the weak (WEC) and null energy conditions (NEC). The first implies $\sigma \geq 0$ and $\sigma + \mathcal{P}\geq 0$, which by continuity implies the second one $\sigma + \mathcal{P}\geq 0$. We provide contour plots for $\sigma$ and $\sigma + \mathcal{P}$ in \cref{fig:3DWEC,fig:3DNEC}, for the fixed value $\beta = 0.1$ and reasonable values of the parameters $y_0$ and $r_0$.
As it can be seen from the plots, both the WEC and NEC are satisfied in these intervals of parameters.

\section{Scalar lumps  with AdS core}
\label{sec:ScalarLump}

Our nonexistence theorem, proved in \cref{sec:NoGodS}, forbids any smooth solution interpolating between a dS spacetime in the core and spacetimes with flat or AdS asymptotics. However, it leaves open the possibility of having solutions interpolating between, for instance, an AdS vacuum in the core and a Minkowski or dS vacua at spatial infinity. We have already discussed the former case in \cref{sec:smoothAdScore} (see also Ref.~\cite{Franzin:2018pwz}). In this section, we will consider an example in which the solution reduces to a dS$_2\times S_2$ Nariai spacetime \cite{nariai1999new} at asymptotic infinity. This represents a regular lump in a nonasymptotically-flat spacetime (see Refs.~\cite{Lavrelashvili:2021rxw,Lavrelashvili:2021qtg} for numerical solutions of this kind).
These solutions are of interest because they could   realize a dynamically phase transition between two different vacua (dS and AdS), leading to a dynamical change of sign of the cosmological constant. Until now, only solutions in which this transition is realized in a nondynamical way \cite{Biasi:2022ktq,Garriga:2013cix} are known.

\begin{figure}
\centering
\includegraphics[width= 8 cm, height = 9 cm,keepaspectratio]{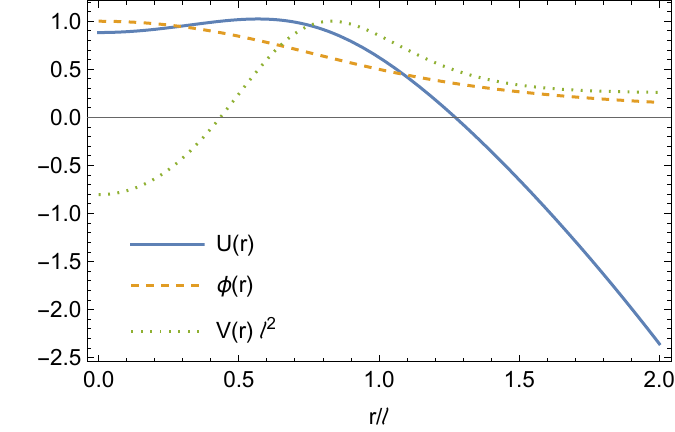} 
\caption{Metric (blue solid line), scalar field (dashed orange line) and scalar potential (dotted green line) as functions of $r/\ell$ for our  scalar lumps. We set $c_2 = \ell^{-1}$ and rescaled each curve by their maximum.}
\label{fig:UphiVLump}
\end{figure}

We start from the radial function 
\begin{equation}
R(r) = \frac{r}{\left[1+\left(\frac{r}{\ell} \right)^4 \right]^{1/4}}\, ,
\end{equation}
where $\ell$ is the usual length-scale parameter. Notice that, near $r\sim 0$, $R(r) \sim r$, while at infinity, $R(r) \sim \ell = \text{constant}$. Therefore, the geometry of all these metrics will reduce asymptotically to $\mathcal{M}^2 \times S^2$, typical of lumps. The function $P(r)$ is
\begin{equation}
P(r) = \frac{\ell}{\left(r^4 + \ell^4 \right)^{1/4}}\, .
\end{equation}
Near $r \sim 0$, $P(r) \sim 1 - r^4/(4 \ell^4)$, which allows, as we shall soon show, for an AdS regular interior. The integration of \cref{Ugeneral} is straightforward. After setting to zero the integration constant $c_1$, necessary to satisfy the regularity conditions, the metric function reduces to
\begin{equation}
U(r) = \frac{c_2 r^2 \ell ^4-r^4+\ell ^4}{\ell ^2 \sqrt{r^4+\ell ^4}}\, , \label{fullsolution}
\end{equation}
which, near $r \sim 0$, behaves as
\begin{equation}
U(r) \simeq 1 + c_2 r^2 - \frac{3 r^4}{2\ell^4} + \mathcal{O}(r^6)\, .
\end{equation}
$|c2|$ determines  the inverse of the square of the (A)dS length in the core, whereas its  sign dictates  the behavior of the metric in  the interior. If $c_2 >0$, the solution exhibits an AdS core (written in global coordinates), while $c_2 < 0$ implies a dS core. $c_2=0$, instead, corresponds to a Minkowski core.  
The sign of $c_2$ is fixed by requiring the signature of the metric to remain the same.  
In fact, as $r\to \infty$ we have 
\begin{equation}\label{Ulump}
U(r) \simeq c_2 \ell^2 - \frac{r^2}{\ell^2} + \mathcal{O}(r^{-2})\, ,
\end{equation}
which forces  $c_2>0$.
In this latter case, a rescaling of the time and radial coordinate, $\tilde t \equiv \sqrt{c_2}\ell \, t$ and $\tilde r \equiv r/(\sqrt{c_2} \ell)$, 
brings the asymptotic metric to the form 
\begin{equation}
\dd s^2 = -\left(1-\frac{\tilde r^2}{\ell^2} \right)\dd \tilde t^2 + \frac{\dd \tilde r^2}{1-\frac{\tilde r^2}{\ell^2}}+\ell^2 d\Omega^2,
\end{equation}
which describes a dS$_2\times S_2$ spacetime with $\ell$ playing the role of both of the dS length and radius of the two-sphere, namely the so-called Nariai spacetime \cite{nariai1999new}. The full solution \eqref{fullsolution}, thus, interpolates between an AdS$_4$  in the UV ($r\to0$) and dS$_2 \times$ S$^2$ in the IR ($r \to \infty$). We plot the metric function $U$ in \cref{fig:UphiVLump} (solid blue line).

We note that, owing to the dS asymptotic behavior, the metric function has a (cosmological) horizon located at $U(\rH) = 0$, which is satisfied by
\begin{equation}\label{rHell}
\rH = \frac{\ell}{\sqrt{2}}\sqrt{c_2 \ell^2 + \sqrt{c_2^2 \ell^4 + 4}}\, .
\end{equation}
The scalar field solution reads as
\begin{equation}\label{philump}
\phi(r) = \sqrt{5} \, \tan^{-1}\left(\frac{\ell^2}{r^2} \right)\, ,
\end{equation}
so near $r \sim 0$, $\phi \simeq \sqrt{5}\pi/2-\sqrt{5}r^2/\ell^2 + \mathcal{O}(r^{-6})$, while at infinity it decays as $r^{-2}$. The scalar-field profile is shown in \cref{fig:UphiVLump} (dashed orange line).

Finally, the potential reads
\begin{equation}\label{Vlump}
V(r) = \frac{2 \left(5 c_2 r^4 \ell ^8-3 c_2 \ell ^{12}+r^{10}+15 r^2 \ell ^8\right)}{\ell ^2 \left(r^4+\ell ^4\right)^{5/2}}\, .
\end{equation}
We show the behavior of $V$ in \cref{fig:UphiVLump} (dotted green line).
The negative minimum at $r = 0$ gives the AdS behavior in the core. Here, the potential behaves as
\begin{equation}\label{eq:mau3}
	V(r) \simeq -6 c_2 + \frac{30 r^2}{\ell^4} + \mathcal{O}(r^4)\, .
\end{equation}
At infinity, instead, $V$ tends to a positive constant, \emph{i.e.}, $V(r) \simeq 2/\ell^2 -5 \ell^2/r^4+ \mathcal{O}(r^{-5})$, which confirms its relation with the dS$_2$ length. 

\section{Summary and conclusions}

\label{sec:conclusions}

Owing to their simplicity, scalar fields have been widely employed in gravitational physics. In cosmology they have been used to describe inflation and dark energy, whereas in the AdS/CFT framework they have been employed to generate holographic phase transitions.  Until now, their use as sources for black-hole mimickers has basically been limited to the case of boson stars, for which the scalar field is complex. This is because very few solutions of Einstein-scalar gravity, describing compact, horizonless, singularity-free, spherically-symmetric objects sourced by a real scalar field, are known. The same is also true for general statements on the existence of these kind of solutions. Classical no hair and nonexistence theorems have been formulated for singular and nonsingular black holes, but not for compact, horizonless objects.

The main obstruction to finding solutions and formulating nonexistence theorems is due to the fact that, in the usual approach, one has to first fix the self-interaction potential $V(\phi)$, drastically restricting the space of possible solutions of the theory.
In this paper, we have overcome this difficulty by parametrizing the solutions in terms of the radial metric function instead of the scalar-field potential. This alternative formulation allowed us to discuss, in a systematic way, a particularly interesting subspace of the full solution space, describing compact objects interpolating between a singularity-free core (in the form of a (A)dS spacetime) and an asymptotically flat spacetime. 
 
We have classified the approximate solutions in the dS core and in the asymptotically-flat region in terms of the allowed subleading terms. The investigation of these asymptotic solutions motivated the proposal of a new class of compact objects that we have termed gravistars. These solutions interpolate between an inner core that is dominated by the potential, and an outer region, dominated instead by the scalar-field kinetic term. We have then proved a nonexistence theorem for smooth solutions and derived explicitly three different classes of exact (smooth and nonsmooth) singularity-free, solutions: (1) smooth solutions interpolating between an AdS spacetime in the core and an asymptotically-flat spacetime;(2) nonsmooth gravistar solutions; (3) smooth scalar lump solutions interpolating between $\text{AdS}_4$ in the core and a Nariai spacetime at infinity.  
 
While our initial aim of constructing regular models with a dS core in Einstein's gravity coupled to a scalar field has proven to be generally unfeasible, with the exception of a limited class of nonsmooth solutions, the other regular configurations we have explored offer viable alternatives to traditional GR compact objects. The deviations they introduce could be tested in the near future, potentially imposing new constraints on the presence of scalar fields in our universe. However, it is crucial to investigate their formation mechanisms and stability in future research.

\newpage

\begin{appendix}
\section{Components of the extrinsic curvature and surface density and pressure of the shell}
\label{AppA:ExtrinsicCurvature}

Using \cref{Extrinsiccurvaturegeneral}, we see that for a static and spherically-symmetric spacetime, the only nonzero components are $K^{\pm}_{tt}$, $K^{\pm}_{\theta\theta}$ and $K^{\pm}_{\varphi\varphi}$. Their form is as follows (prime always refers to derivation with respect to $r$)

\begin{equation}\begin{split}
K^+_{tt} & = -\Gamma^{(+) r}_{tt} n^+_r = -\frac{1}{2}\sqrt{U^+(r)}\, \left(U^+(r)\right)'  = \frac{2\beta-1}{2r^2}\, r_0 \, \left(1-\frac{r_0}{r} \right)^{\frac{1}{2}-3\beta}\, .
\end{split}
\end{equation} 

\begin{equation}\begin{split}
K^-_{tt} & = -\Gamma^{(-) r}_{tt} n^-_r \\
& = \frac{1}{3\ell^6 \left(r^4+\ell^4 \right)^{5/2}} \sqrt{-2\ell^4 r^4 - \frac{r^8}{3}+\ell^8\left(1+c_2 r^2 \right)}\, \left[3c_2 \ell^8 r \left(r^4-\ell^4 \right)+2r^3\left(r^8 + 9\ell^8 + 2\ell^4 r^4 \right) \right]\, .
\end{split}
\end{equation}

\begin{equation}
\begin{split}
K^+_{\theta\theta} & =-\Gamma^{(+) r}_{\theta\theta} n^+_r = \sqrt{U^+(r)}\,  R^+(r) \left(R^+(r) \right)' = \frac{r}{r-r_0}\left(1-\frac{r_0}{r} \right)^{\beta + \frac{1}{2}} \, \left[r + r_0\left(\beta-1\right) \right]\, .
\end{split}
\end{equation}

\begin{equation}
\begin{split}
K^-_{\theta\theta} & =-\Gamma^{(-) r}_{\theta\theta} n^-_r  = \sqrt{U^-(r)}\,  R^-(r) \left(R^-(r) \right)' = \frac{\ell^2 r \left(\ell^4-r^4 \right)}{\left(r^4 + \ell^4 \right)^{5/2}} \, \sqrt{-2\ell^4 r^4 - \frac{r^8}{3}+\ell^8\left(1+c_2 r^2 \right)}\, .
\end{split}
\end{equation}

\begin{equation}
K^{\pm}_{\varphi \varphi} = \sin^2 \theta K^{\pm}_{\theta\theta}\, .
\end{equation}
We also verified that the quantity under the square roots is indeed positive for the values of the constants considered in \cref{subsec:continuitycondition}.
Then, the traces of the extrinsic curvatures inside and outside the shell are
\begin{equation}
K^+ = h^{(+)ij}K^+_{ij} = \left(1-\frac{r_0}{r} \right)^{\frac{1-2\beta}{2}} \, \frac{4r + r_0\left(2\beta-3 \right)}{2r\left(r-r_0 \right)} \, ;
\end{equation}

\begin{equation}
K^- = h^{(-)ij}K^-_{ij} = \frac{\ell^2 \left[2r^8-3\ell^4r^4 \left(4 + c_2 r^2 \right) + \ell^8 \left(2+3c_2 r^2 \right) \right]}{r\left(\ell^4+r^4 \right)^{3/2} \sqrt{-2\ell^4 r^4 - \frac{r^8}{3}+\ell^8 \left(1+c_2 r^2 \right)}}\, .
\end{equation}

Finally, using \cref{SETshellcomponents}, the explicit expressions of the surface density $\sigma(y_0)$ and pressure $ \mathcal{P}(y_0) $ of the shell are 
\begin{equation}\begin{split}
\sigma(y_0) = &-\frac{1}{8 \pi  y_0^2}\left(1-\frac{r_0}{y_0}\right)^{-2 \beta} \biggl[\frac{\ell ^2 \left(\ell ^8 \left(3 c_2 y_0^2+2\right)-3 y_0^4 \ell ^4 \left(c_2 y_0^2+4\right)+2 y_0^8\right)}{y_0 \left(y_0^4+\ell ^4\right){}^{3/2} \sqrt{c_2 y_0^2 \ell ^8-2 y_0^4 \ell ^4-\frac{y_0^8}{3}+\ell ^8}}+\\
&-\frac{y_0 \ell ^2 \left(\ell ^4-y_0^4\right) \sqrt{c_2 y_0^2 \ell ^8-2 y_0^4 \ell ^4-\frac{y_0^8}{3}+\ell ^8}}{\left(y_0^4+\ell ^4\right)^{5/2}}+\frac{y_0 \sqrt{\left(1-\frac{r_0}{y_0}\right)^{1-2 \beta }} \left((\beta -1) r_0+y_0\right) \left(1-\frac{r_0}{y_0}\right){}^{2 \beta }}{y_0-r_0}+\\
&-\frac{\sqrt{\left(1-\frac{r_0}{y_0}\right){}^{1-2 \beta }} \left((2 \beta -3) r_0+4 y_0\right)}{2 y_0 \left(y_0-r_0\right)}\biggr]\, ;
\end{split}
\label{sigmay0}
\end{equation}

\begin{equation}
\begin{split}
\mathcal{P}(y_0) =& -\frac{1}{8 \pi  y_0^2}\left(1-\frac{r_0}{y_0}\right)^{-2 \beta } \biggl[\frac{\ell ^2 \left(\ell ^8 \left(3 c_2 y_0^2+2\right)-3 y_0^4 \ell ^4 \left(c_2 y_0^2+4\right)+2 y_0^8\right)}{y_0 \left(y_0^4+\ell ^4\right){}^{3/2} \sqrt{\ell ^8 \left(c_2 y_0^2+1\right)-2 y_0^4 \ell ^4-\frac{y_0^8}{3}}}+\\
&-\frac{y_0 \ell ^2 \left(\ell ^4-y_0^4\right) \sqrt{\ell ^8 \left(c_2 y_0^2+1\right)-2 y_0^4 \ell ^4-\frac{y_0^8}{3}}}{\left(y_0^4+\ell ^4\right){}^{5/2}}+\frac{y_0 \sqrt{\left(1-\frac{r_0}{y_0}\right){}^{1-2 \beta }} \left((\beta -1) r_0+y_0\right) \left(1-\frac{r_0}{y_0}\right){}^{2 \beta }}{y_0-r_0}+\\
&-\frac{\sqrt{\left(1-\frac{r_0}{y_0}\right){}^{1-2 \beta }} \left((2 \beta -3) r_0+4 y_0\right)}{2 y_0 \left(y_0-r_0\right)}\biggr]\, .
\end{split}
\end{equation}

\end{appendix}

\newpage

\bibliography{ScalarFieldRegularFinal}
\bibliographystyle{ieeetr}

\end{document}